\documentclass[aps,prd,nofootinbib]{revtex4}
\usepackage{graphicx}
\usepackage{amsmath,amssymb,latexsym}

\begin{document}
\title{Towards the Final Fate of an Unstable Black String}

\author{Matthew Choptuik$^{1,2}$, Luis Lehner$^{3,2}$, Ignacio (I\~naki) Olabarrieta$^{2}$, \\
        Roman Petryk$^{2}$, Frans Pretorius$^{4,2}$ and Hugo Villegas$^{2}$}

\address{$^{1}$ CIAR Cosmology and Gravity Program \\
         $^{2}$ Department of Physics and Astronomy\\
         University of British Columbia,\\
         Vancouver, CANADA V6T 1Z1 \\
         $^{3}$ Department of Physics and Astronomy\\
         Louisiana State University, \\
         Baton Rouge, LA 70810.\\
         $^{4}$ Theoretical Astrophysics 130-033\\
         California Institute of Technology,\\
         Pasadena, CA, 91125}
\begin{abstract}
Black strings, one class of higher dimensional analogues of black
holes, were shown to be unstable to long wavelength perturbations
by Gregory and Laflamme in 1992, via a linear analysis.
We revisit the problem through numerical solution of the full equations
of motion, and focus on trying to determine the
end-state of a perturbed, unstable black string. Our preliminary
results show that such a spacetime tends towards a solution
resembling a sequence of spherical black holes connected by thin
black strings, at least at intermediate times. However, our code
fails then, primarily due to large gradients that develop
in metric functions, as the coordinate system we use is not
well adapted to the nature of the unfolding solution. We are thus
unable to determine how close the solution we see is to the final
end-state, though we do observe rich dynamical behavior of the system
in the intermediate stages.
\end{abstract}
\maketitle

\section{Introduction}
The stability of four-dimensional black holes is a well known,
and fundamental result of relativity theory~\cite{chandra}.
The picture in higher dimensional
spacetimes was shown to be quite different by Gregory and 
Laflamme~\cite{gregorylaflame1,gregorylaflame2}, who demonstrated 
the existence of unstable long wavelength modes of
the black string in perturbation theory. 
This finding, coupled to arguments based on entropy considerations, led
to their conjecture that black strings might bifurcate into higher
dimensional analogues of spherical black holes. 
Cosmic censorship would be violated were this the case,
since a singularity must be
encountered by a bifurcating black hole horizon, essentially as a 
consequence of the principle of equivalence~\cite{hawkingellis}.

The existence of the Gregory-Laflamme instability has been assumed in 
many subsequent studies of higher dimensional gravity theory, including 
the classical limit of string theory 
(see, for example, \cite{hms96,imsy98,agmoo00,hv00}), some of which 
have also assumed the validity of the bifurcation conjecture.
However, a linearized analysis can say little, if anything, concerning 
the nature of the full non-linear evolution of an unstable string, and 
the final end-state of such a configuration remained to be established.

Recently,  Horowitz and Maeda were able to prove, under some
assumptions, that black strings cannot bifurcate in finite
time~\cite{horowitzmaeda}. Furthermore, they conjectured that the
system is likely to approach a new stationary solution which is
not translationally invariant along the string direction. However, even if
the assumptions involved in the proof are sufficiently generic, their
analysis cannot identify the final end-state of evolution. 
Partial answers can be
sought via perturbation analysis as done by Gubser~\cite{gubser}.
By assuming the Horowitz-Maeda conjecture and linearizing
the solution {\em at the critical length} to first order (and a
partial extension to second order), Gubser argued that the transition
to the final solution must be of second order type (i.e.
discontinuous). Despite these developments, it seems clear that a convincing 
answer to the question at hand
can only be obtained by solving the full equations governing the
problem. A step in this direction would be to search for special
solutions, such as stationary ones, and compare the physical content of the
obtained configurations with the black string solutions. This has recently
been carried out by Wiseman~\cite{wiseman}, who numerically
solves the equations resulting from a static ansatz.
Interestingly, he finds non-uniform solutions with mass larger than
that of the black string for a given compactification radius.
Thus, he concludes that the solutions he finds
cannot be the end-states
conjectured by Horowitz and Maeda (also see related work by Kol~\cite{kol}).

Additional work by Unruh and Wald~\cite{unruhwald} studies the dynamics of a
uniform cylindrical configuration of matter in Newtonian gravity. They observe
that a perturbation of the density gives raise to a Jeans instability 
 responsible for the collapse of the system along the cylinder's length.
They then argue that if the main features of this model are robust in the passage to the
general relativistic system, one possible end-state for the perturbed black
string would be collapse in the string direction, resulting in singularity
formation.
Note that this collapse need not lead to violations
of cosmic censorship, as the final
singularity could still be hidden by an event horizon~\cite{geddes}.

Clearly there are several distinct viable possibilities for the final 
end-state of a perturbed black string, with 
remarkably different
consequences associated with the range of options.
Current conjectures range from ``nothing interesting happens'', to
violations of cosmic censorship, to the arguably more extreme case of
a complete collapse of the
spacetime. In order to completely settle the issue,
the full dynamics of the perturbed black string needs to be addressed. 
At least in principle, this will allow us to identify which of the 
above possibilities (if any!) is actually realized. 
In this paper we report on preliminary work in this
direction---a program to simulate the dynamics of the black string
through numerical solution of the Einstein equations. 
At this stage of the
project, we cannot yet provide an answer to the question of the
end-state; however, we have tantalizing
results that show the spacetime going through a configuration
resembling 5-dimensional spherical black holes connected by thin
black strings that expand along the string dimension. Our simulations
eventually crash while the spacetime is still fairly dynamical,
and so we cannot determine whether what we see is near the end-state, or
merely an intermediate configuration in a more complicated
evolution. Underlying the current failure of our simulations is the
fact that the coordinate system employed is not well adapted to the solution
that unfolds at late times, wherein fatally steep gradients develop in
metric functions~\footnote{This divergence of metric gradients 
does not happen earlier as
resolution is increased, and is not accompanied by divergence of curvature
invariants such as the Kretschmann scalar. This suggests that the code is 
evolving to a coordinate, rather than a geometric, singularity.}.

The outline of the remainder of the paper is as follows.
In Sec.~\ref{sec:setup},
we begin by describing the equations of motion, our
coordinate choices, generation of initial data, as well as our numerical solution scheme.
Additionally, we also briefly mention the tools
we employ to monitor the solution, deferring details to the Appendices.
In Sec. \ref{sec:results} we discuss the results obtained with this code, and
conclude in Sec. \ref{sec:conclusion} by mentioning directions for future work
that may allow us to more definitively answer questions regarding the
end-state of the Gregory-Laflamme instability.

\section{Equations, Black Strings and  Numerics}\label{sec:setup}
We wish to solve the vacuum Einstein equations in higher dimensional settings.
For simplicity, and without loss of
generality in studying the Gregory-Laflamme instability, we only consider
the 5-dimensional case, and restrict attention
to spherical symmetry within the 4-dimensional subspace tangent
to the ``extra'' dimension. We also use the natural generalization of the 
ADM decomposition to derive the system of equations that we 
then solve numerically.
Choosing units in which $G=c=1$, and adopting MTW~\cite{MTW} conventions,
our starting point is thus a metric element given by
\begin{eqnarray}\label{metric}
ds^2=(-\alpha^2+\gamma_{AB} \beta^A \beta^B) dt^2 + 2 \gamma_{AB} \beta^A dx^B
dt + \gamma_{AB} dx^A dx^B + \gamma_{\Omega} d\Omega^2
\end{eqnarray}
where $x^A=(r,z)$, and $d\Omega^2$ is the 2-spherical line element
with coordinates chosen orthogonal to the $t=$ constant
congruences (hence there is no shift corresponding to angular
directions). All metric components defined via (\ref{metric}) depend upon
$(t,r,z)$: $t$ is a time-like coordinate, $r$ is a radial coordinate, 
and $z$ is the coordinate along the length of the string.  To further 
expedite the numerical implementation,
we make $z$ a periodic coordinate
by identifying $z=0$ and $z=L$. Then, following the results
of Gregory and Laflamme, we expect black strings
to be unstable only for $L$ greater than some critical length $L_c$.

The vacuum Einstein equations, written in ADM form \cite{MTW,york_78},
are 1) the Hamiltonian constraint
\begin{equation}\label{HC}
H \equiv ^{(4)}R + K^2 - K_{ab} K^{ab} = 0,
\end{equation}
2) the momentum constraints
\begin{equation}\label{MC}
M_a \equiv K_{a\ \ |b}^{\ b} - K_{|a} = 0,
\end{equation}
3) the evolution equations for the $\gamma_{ab}$
\begin{equation}\label{gdot}
\frac{\partial \gamma_{ab}}{\partial t} = - 2 \alpha K_{ab}
+ \beta_{a|b} + \beta_{b|a},
\end{equation}
that follow from the definition of the extrinsic curvature 
$K_{ab}$ associated with $t={\rm const.}$ slices,
and 4) the evolution equations for the extrinsic curvature
\begin{eqnarray}\label{kdot}
\frac{\partial K_{ab}}{\partial t} &=& \alpha \left(^{(4)}R_{ab} +
K K_{ab}\right) -2\alpha K_{a c} K^c{}_{b} -\alpha_{|ab} \nonumber
\\& & + \beta^c{}_{|a} K_{cb} + \beta^c{}_{|b} K_{ca} +\beta^c K_{ab|c}
+ \alpha F^c_a F^d_b \gamma_{cd}  H .
\end{eqnarray}
In the above, $a,b,\ldots$ are four-dimensional (spatial) indices, 
$^{(4)}R_{ab}$ and $^{(4)}R$ are, respectively, the
Ricci tensor and Ricci scalar intrinsic to the four-dimensional spatial
hypersurfaces, $\alpha$ is the lapse function, $\beta^c$ is 
the shift vector, the vertical bar $_|$ denotes covariant
differentiation in the spatial hypersurfaces (compatible with $\gamma_{ab}$),
and $F^c_a = -2 \delta^c_r \delta^r_a$.  We note that the term proportional 
to $H$ in~(\ref{kdot}) has been added as a result of stability considerations;
see for instance, the discussions in~\cite{lehnerreview,shinkai,kelly}. 
To simplify the final set of 
equations solved numerically, as well as to regularize certain terms that 
otherwise diverge
at spatial infinity (see the discussion of our coordinate system
in the next sub-section), we define the following variables
\[
g_{rr} \equiv \gamma_{rr}  \quad\quad\quad g_{rz} \equiv \gamma_{rz} 
	\quad\quad\quad g_{zz} \equiv \gamma_{zz}
\]
\[
g_{\theta\theta} \equiv \gamma_{\Omega}/r^2  
	\quad\quad\quad g_{\phi\phi} \equiv \gamma_{\Omega}/(r^2\sin^2\theta)
\]
\[
k_{rr} \equiv r^2 K_{rr}/\alpha  \quad\quad\quad k_{rz} \equiv K_{rz} 
	\quad\quad\quad k_{zz} \equiv K_{zz}
\]
\begin{equation}
k_{\theta\theta} \equiv K_{\theta\theta}/\alpha
	\quad\quad\quad k_{\phi\phi} \equiv K_{\theta\theta} / (\alpha \sin^2\theta)
\end{equation}
and use them as the fundamental dynamical quantities in our numerical code.
As discussed in the following sub-sections,
to complete the prescription of the evolution problem 
we need to choose a suitable lapse and shift, specify
initial and boundary conditions, and then implement these choices 
and specifications consistently.

\subsection{Boundary and Coordinate Conditions}
A particular concern here is that ``standard'' outer boundary
conditions~\cite{lehnerreview}, often imposed during numerical
evolution of Einstein's equations, 
might not be well suited for studying the string
instability. In particular, we must
be able to evolve for very long times while absolutely minimizing spurious
influences from the outer boundary of the computational domain. 
In addition, in the present case
we cannot assume, {\em a priori}, that any given initial configuration will 
settle down to some stationary solution; thus, boundary conditions predicated
on such assumptions (such as a $1/r$ fall-off condition in a metric
component), when imposed at a finite proper
distance from the black string, could very well adversely affect the 
numerical results~\footnote{In fact, in an earlier version of the code that did
not use a radially compactified coordinate system, we did
encounter such problems, in that some artificial stationary
non-homogeneous solution
was apparently entirely ``sourced'' via an outer boundary located at 
a finite distance from the string.}. To
ensure minimal boundary influence we therefore extend the domain
of integration to $i^o$ by radially compactifying the spacelike
hypersurfaces via the introduction of a new coordinate, $x$, defined by
\begin{equation}
x\equiv \frac{r}{1+r} \, .  
\label{defx}
\end{equation}
As might be expected, this transformation causes computational problems of 
its own---most notably decreased spatial resolution at large distances---but, 
as discussed in Sec.~\ref{numerics}, we can deal with these 
difficulties using numerical dissipation.
Having introduced the new compactifying coordinate, we can directly impose 
boundary conditions derived from the demand of asymptotic flatness at 
spatial infinity, which lies at $x=1$.

We employ singularity excision techniques~\cite{unruh} to allow us
to evolve the entire perturbed black string spacetime exterior to
the apparent horizon (plus a certain ``buffer zone'' that lies within
the horizon).
Hence, we do not need to impose inner boundary conditions as long as the 
$t=\rm{const.}$ hypersurfaces penetrate the horizon, and that all 
characteristics of the evolution equations are in-going on the 
boundary. 
Ensuring that this is the case involves choosing ``good''
coordinate conditions (choice of lapse and shift), which, for generic 
string evolutions,  remains an open problem.
As a preliminary step, we have based our coordinate choices on 
those that yield the ingoing Eddington-Finkelstein form
of the unperturbed black string metric
\begin{equation}\label{bs_metric}
ds^2_{\rm BS}= -(1-2M/r) dt^2 + 4 M/r dr dt + (1+2M/r) dr^2 + dz^2 +
r^2 d\Omega^2 \, .
\end{equation}
Comparison with the general 5-dimensional ADM form provides the 
identifications 
\begin{eqnarray}
\alpha_{\rm BS} &=& (1+2M/r)^{(-1/2)}\, ,  \label{bs_lapse}\\
\beta_{\rm BS}^A &=& (2M/(r+2M))\delta^A_r\,.  \label{bs_shift}
\end{eqnarray}
For reference we
also list the two non-trivial components of the
extrinsic curvature of a $t={\rm const.}$ slice defined by (\ref{bs_metric}):
\begin{eqnarray}\label{bs_K}
K_{r r} &=& -2M \frac{(r+M)}{r^3} \sqrt{\frac{r}{r+2M}}, \nonumber \\
K_{\theta \theta} &=& 2M \sqrt{\frac{r}{r+2M}}.
\end{eqnarray}
In generalizing~(\ref{bs_lapse}) and (\ref{bs_shift}) to the dynamical case,
we have chosen $\alpha=\alpha_{\rm BS}$ and $\beta^{z}=\beta_{\rm BS}^{z}=0$. In a
preliminary version of our code, we also required that
$\beta^{r}=\beta_{\rm BS}^{r}$. This, however, caused a coordinate
pathology to develop at late times during the evolution of
unstable strings---specifically, some regions of the horizons 
approached a zero coordinate-radius, while maintaining {\em finite}
proper radius. 
In our current efforts, we
choose $\beta^{r}$ such that $g_{\theta \theta}$ remains constant
during evolution \cite{seidelexcision,lehnercadm}, by requiring that
\begin{equation}\label{br_gauge}
\beta^r = \frac{2 \alpha K_{\theta \theta}}{{\gamma_{\theta \theta}}_{,r}}.
\end{equation}
This shift condition performs reasonably well, as will
be seen in Sec. \ref{sec:results}. However, our current simulations still
suffer from ``grid-stretching'' problems in the $z$ direction at late times,
suggesting that a more dynamical gauge condition for $\beta^z$
(and possibly for $\alpha$ and $\beta^{r}$) could be useful.  This issue 
is discussed in more detail in Sec.~\ref{sec:conclusion}.

\subsection{Initial Data \label{ID}}
As anticipated, 
we observe that even numerical truncation errors,
if non-uniform in the $z$ direction, are enough to 
trigger the Gregory-Laflamme instability in our simulations. 
However, to reduce the computational effort required to reach the 
``interesting'' (i.e. non-perturbative) stages of
evolution, we adopt initial configurations whose departure from
the black string solution can be arbitrarily tuned. In order to
find such data, we must solve the Hamiltonian constraint, and the $r$
and $z$ components of the momentum constraint (the other 
components of the momentum constraint are trivially satisfied 
because our coordinate system is adapted to spherical symmetry). 
The deviation---not necessarily 
small---from the black string solution, is introduced 
via $g_{\theta \theta}$, and takes the following form:
\begin{equation}
\label{gtt}
g_{\theta \theta}(0,r,z) = 1 + A\sin{\left(z \frac{2 \pi q}{L}\right)}
                          e^{ -(r-r_o)^2 / \delta_r^2 }.
\end{equation}
Here, $A$ is used to control the overall strength of the ``perturbation'',
while $q$ is an
integer that controls the spatial frequency in the $z$ direction.
For the results presented below, $A=0.1$, $q=1$, $r_0=2.5$ and
$\delta_r=0.5$, and we perturb about a unit mass ($M=1$)
black string solution.
As described in more detail in Appendix \ref{app:initialdata},
$g_{rr}$, $k_{rr}$, $k_{\theta \theta}$ are then calculated by
solving the constraint equations, with the remainder
of the metric and extrinsic curvature variables set
to the values they would take for an unperturbed black string
(see (\ref{bs_metric}) and (\ref{bs_K})).

\subsection{Numerical Evolution\label{numerics}}

To numerically evolve the initial data sets described above,
we discretize the evolution equations~(\ref{gdot})-(\ref{kdot}) using 
second-order accurate finite difference techniques that include 
Crank-Nicholson treatment of the temporal and spatial derivatives.
We use a uniform distribution of grid points in $z$ and $x$
(recall that $x\equiv r/(1+r)$). 
The resulting implicit system of algebraic equations is solved 
iteratively.
We initially implemented a serial version of the algorithm, and later 
coded a parallel version using the CACTUS Computational Toolkit~\cite{cactus},
wherein the equations of motion, monitoring tools and I/O were handled by our
own routines, suitably interfaced to CACTUS.

Black hole excision is handled as follows. We periodically find
the apparent horizon, as discussed in the following
sub-section and Appendix \ref{app:AH}. We then define the surface
along which we excise to be a certain number of ``buffer'' points
{\em inside} the apparent horizon 
(typically $10$---$30$ buffer points are used)~\footnote{In several
tests, we also adopted an excision region given by the global minimum
$r$ value of the apparent horizon and compared the results with those obtained
when the excision region was defined by the apparent horizon. 
The agreement obtained
gives extra indication that the excision implementation is consistent.}.
During each Crank-Nicholson iteration, all the evolution
difference equations are applied up to the excision surface, and
any function values referenced by finite difference stencils
interior to this surface are defined via fourth order
extrapolation. When the apparent horizon location, and hence excision surface
changes during evolution, function values at all repopulated
points (i.e. those that moved from inside to outside the excision
surface during the time step) are computed via the same fourth
order extrapolation routine. The one exception to this procedure is for 
the grid values of 
$g_{\theta\theta}$, which we specify {\em a priori} on the
entire computational domain, and that remain fixed due to our gauge
choice (\ref{br_gauge}). Moreover, we have found it useful to
choose a functional form for $g_{\theta\theta}$ that tends to zero
at some positive value of $r$ (though inside the original apparent horizon
location and outside of the limits of integration of the initial
data). 
This causes the ``pinching-off'' of the unstable black
string to be less severe in coordinate space, i.e. we approach
zero areal radius at a finite coordinate $r$.
In turn, this slightly reduces the virulence of the coordinate problems we 
observe at late times, and also provides better
load-balancing of the parallel code, given the method CACTUS uses to
distribute grids among processors. 

For the evolution, we choose a time step $dt=\lambda_{{\rm CFL}} {\rm min}(dr,dz)$, where
the constant $\lambda_{{\rm CFL}}$ must be set less than $1/\sqrt{2}$ in order
to satisfy the Courant-Friedrichs-Lewy (CFL) stability condition that 
results from our iterative solution of the Crank-Nicholson scheme (typically
we use $\lambda_{{\rm CFL}}=0.25$). Note that this restriction on $\lambda_{{\rm CFL}}$ is
based upon flat-space light speeds within our coordinate system ((\ref{bs_metric})
with $M=0$), which, for the solutions presented here, are always greater than
or equal to the actual coordinate light speeds. 
The function ${\rm min}(dr,dz)$
is calculated by only considering mesh spacings within
the non-excised portion of the coordinate domain. Thus, as the excision
surface moves, $dt$ changes with time since our grid is uniform in $x$, and 
hence non-uniform in $r$.

Crucially, 
we add Kreiss-Oliger-style \cite{kreissoliger} numerical
dissipation to the evolution equations to control unphysical
high-frequency solution components (``noise'') that would otherwise
arise during the simulations. This is particularly helpful at the
excision surface, and near $i^0$, where the radial
compactification of points causes all outgoing wave-like
components of variables to eventually become poorly
resolved.  Smoothing of the high frequency components via the 
Kreiss-Oliger dissipation---which only targets wavelengths of size on 
the order of the mesh spacing---prevents them from inducing numerical 
instabilities near the outer boundary. 

\subsubsection{Monitoring the Evolution}\label{sec_monitor}

To elucidate the nature of our computed spacetimes, we monitor the 
following
quantities: 1) the location of the apparent horizon (which is also used 
for excision as explained earlier), 2) the trajectories of null geodesics
that, to a certain extent, should trace the event horizon, and
3) the Kretschmann invariant $I$ (the square of the Riemann tensor):
\begin{eqnarray}\label{I}
I = R_{\alpha\beta\gamma\delta}R^{\alpha\beta\gamma\delta}.
\end{eqnarray}

If cosmic censorship holds---and results from our current simulations
provide no evidence to the contrary---then any apparent horizon found
will always be inside an event horizon. As is well known, although the apparent horizon can
often be used as a reasonable approximation to the event horizon, the 
two do {\em not}, in general, coincide~\footnote{Indeed, depending
on the slicing an apparent horizon need not exist at
all~\cite{waldah}.}. Clearly, the event horizon is the quantity of interest in studying
the Gregory-Laflamme instability, and therefore we would like to
locate it, or at least a reasonably good approximation to it, in 
our simulation results. Such an approximation 
can be obtained by looking for the boundary of the causal
past of some $r={\rm const.}$ surface that is sufficiently far
outside the apparent horizon that it is certain not to be inside the event horizon, yet
close enough to the apparent horizon that its causal past, tracing backwards
from the end-time of the simulation, probes the region of interest
of the spacetime. We use a  method to find the
approximate event horizon discussed by Libson et al. \cite{eh}. The approach is
based on radial outgoing null geodesics; as explained in \cite{eh}, the stable direction
for the integration of null rays that emanate
from the vicinity of an event horizon is backwards in time $t$. Thus, once
we have the complete data from the entire evolution, we start with data 
from the latest time step available, and trace the null rays 
backwards in time.

Monitoring curvature scalars is useful in obtaining coordinate
independent information about a numerical solution of the Einstein 
equations. In particular, $I$, as defined by~(\ref{I}) 
evaluates to $I_{\rm BS}=48M^2/\gamma_{\theta\theta}{}^3$
for the unperturbed black string solution, and as
$\gamma_{\theta\theta}$ is an invariant in spherical symmetry
(i.e. it is proportional to the area of an $r={\rm const.}$
2-sphere), we can compare $I_{\rm BS}$ to the values of $I$ computed from 
a numerical solution to get some indication of how close the computed solution 
is to a  black string spacetime. Furthermore, we can examine $I$ to
see whether curvature singularities (other than the central $r=0$
singularity) may be forming prior to the demise of the simulation that 
invariably occurs when sufficiently steep metric
gradients develop. We note, however, that if $I$ does {\em not} diverge,  it
does not necessarily follow that the geometry is remaining non-singular;
we would need to examine a larger set of curvature
scalars to be certain that the solutions are remaining free of 
physical singularities.

Appendix \ref{app:AH} contains an explanation of the method we used to find
apparent horizons, while
details of the integration techniques aimed at approximately locating event 
horizons can be found in Appendix \ref{app:null}.

\section{Results}\label{sec:results}
In this section we present results from our preliminary study of the 
black string
instability. After briefly showing that we recover some of the key 
Gregory-Laflamme results in
the next sub-section, we present a detailed analysis of a typical unstable
case in Sec. \ref{sec:unstable}. In the following, we will use the value of 
$L_c\approx14.3M$ (with $M=1$) found by Gubser \cite{gubser}, 
which is more accurate than the value we can estimate from the 
zero-crossing of the (positive mode) interpolating curve presented 
in~\cite{gregorylaflame1}.

\subsection{Recovery of Gregory-Laflamme Results}
We ran a variety of simulations of black strings that were  perturbed 
according to the prescription discussed in
section \ref{ID}.  We concentrated on cases with  $L$ ranging from $0.6 L_c$ to $1.8 L_c$
and defined $g_{\theta\theta}$ via equation (\ref{gtt}). 
In general, we observed the expected instability for $L>L_c$, though
for the maximum resolution at which we performed this survey 
($800$ grid points in $r$
and $200$ points in $z$), we could only confirm $L_c$ to within about
$2\%$ of the expected value. In this regard we note that as 
$L$ approaches $L_c$ from
above, the growth rate of the instability goes to $0$, requiring
longer time integrations to identify the instability, which, in turn,
demands ever increasing resolution
to counter the effects of accumulating numerical errors. Furthermore,
the initial configurations we have adopted contain energy in the 
form of gravitational waves, 
and some of this energy falls into the string early on during the evolution.
The increase in the mass of the string (based upon the increase in area of the apparent 
horizon) is typically around $0.3-0.5\%$, and we would have to take this into
account were we to attempt to determine $L_c$ from our simulations 
to a higher degree of accuracy.

\begin{figure}
\begin{center}
\includegraphics[width=15cm,clip=true]{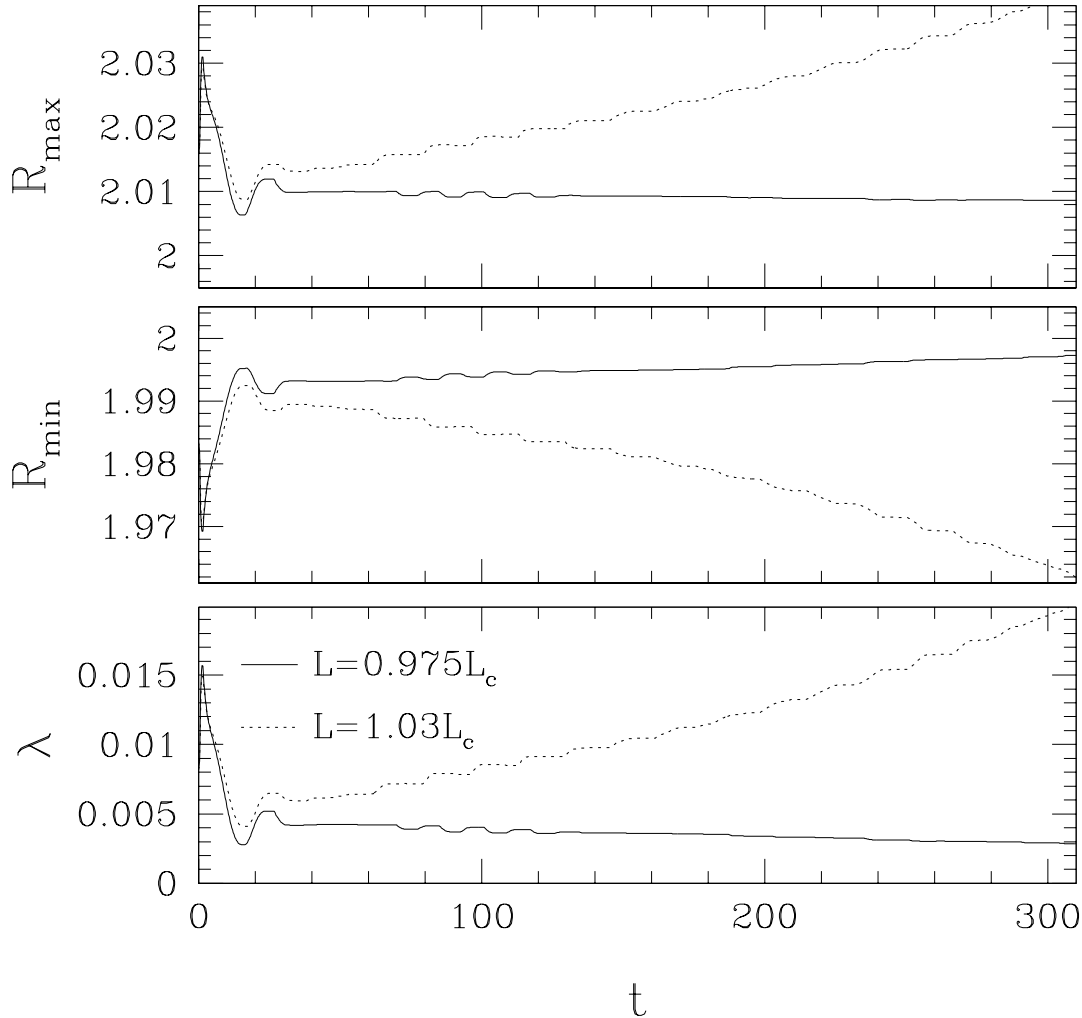}
\end{center}
\caption{The maximum ($R_{{\rm max}}$) and minimum ($R_{{\rm min}}$) areal radii,
and the corresponding function $\lambda$ of the
apparent horizon as a function of time, from the evolution of
perturbed black strings with 
$L=1.03L_c$ and $L=0.975L_c$. The initial
fluctuation in the plots correspond to the effect of the initial
gravitational wave perturbation, most of which either falls into
the string, or escapes to infinity. This close to the threshold $L_c$,
the growth/decay of the remnant perturbation is quite slow, and so
we cannot feasibly (at the resolution of the these simulations---$800 \times 200$ points in $r \times z$) follow the evolution for much
further than shown while maintaining reasonable accuracy (though we see no 
signs of numerical instabilities in the stable case, and such simulations have 
been followed to $10,000M$). 
However, the main purpose
of this figure is to demonstrate the qualitative recovery of
the expected threshold behavior for the onset of the instability at
$L=L_c$.
\label{lambda_comp}}
\end{figure}

As a demonstration of the ability of our code to ``bracket'' the instability, 
and following the notation of~\cite{gubser},
Fig. \ref{lambda_comp} shows a plot of, $\lambda$, defined by
\begin{equation}\label{lambda}
\lambda=\frac{1}{2}\left(\frac{R_{{\rm max}}}{R_{{\rm min}}}-1\right)
\end{equation}
for $L=0.975L_c$ and $L=1.03L_c$.
In the above, $R_{{\rm max}}$ and $R_{{\rm min}}$ are the maximum and minimum 
areal radii, respectively, of the
apparent horizon at some $t={\rm const.}$ slice of the spacetime. 
In particular,  we have $\lambda=0$ for
the static black string spacetime.

\subsection{Beyond the Linear Regime}\label{sec:unstable}
\begin{figure}
\begin{center}
\includegraphics[width=15cm,clip=true]{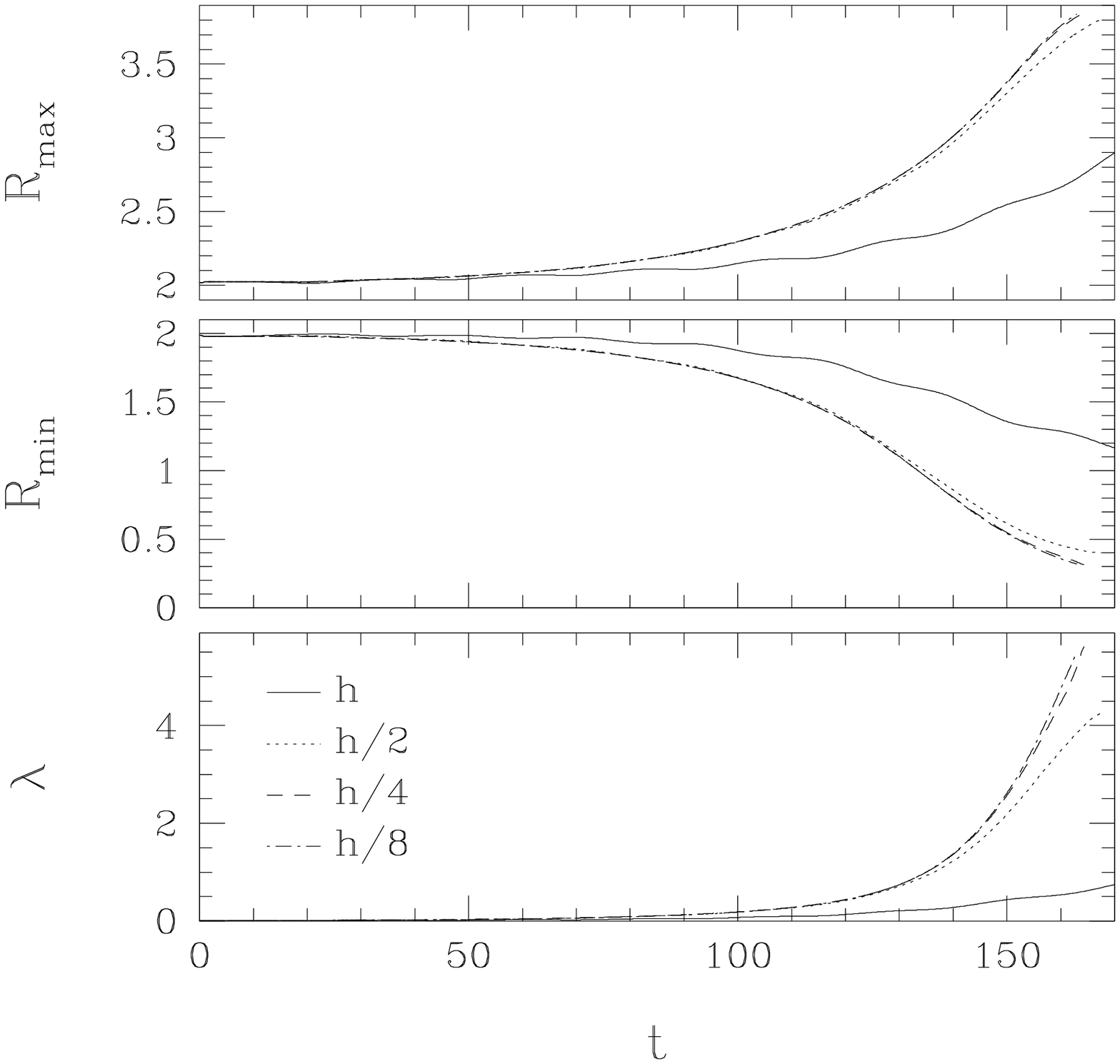}
\end{center}
\caption{The maximum ($R_{{\rm max}}$) and minimum ($R_{{\rm min}}$) areal radii,
and the corresponding function $\lambda = (R_{\rm max}/R_{\rm min} - 1)/2$,
of the apparent horizon, as a function of time, from the evolution of a
perturbed black string with $L=1.4L_c$. 
$h$ labels grid spacing; hence smaller $h$ corresponds to higher
resolution. This plot, combined with the results shown in Fig.~\ref{hc_1000} 
suggest that the code is in the convergent regime---in particular at later 
times---for the $h/2$ and higher resolution 
simulations.
\label{lambda_1000}}
\end{figure}

\begin{figure}
\begin{center}
\includegraphics[width=14cm,clip=true]{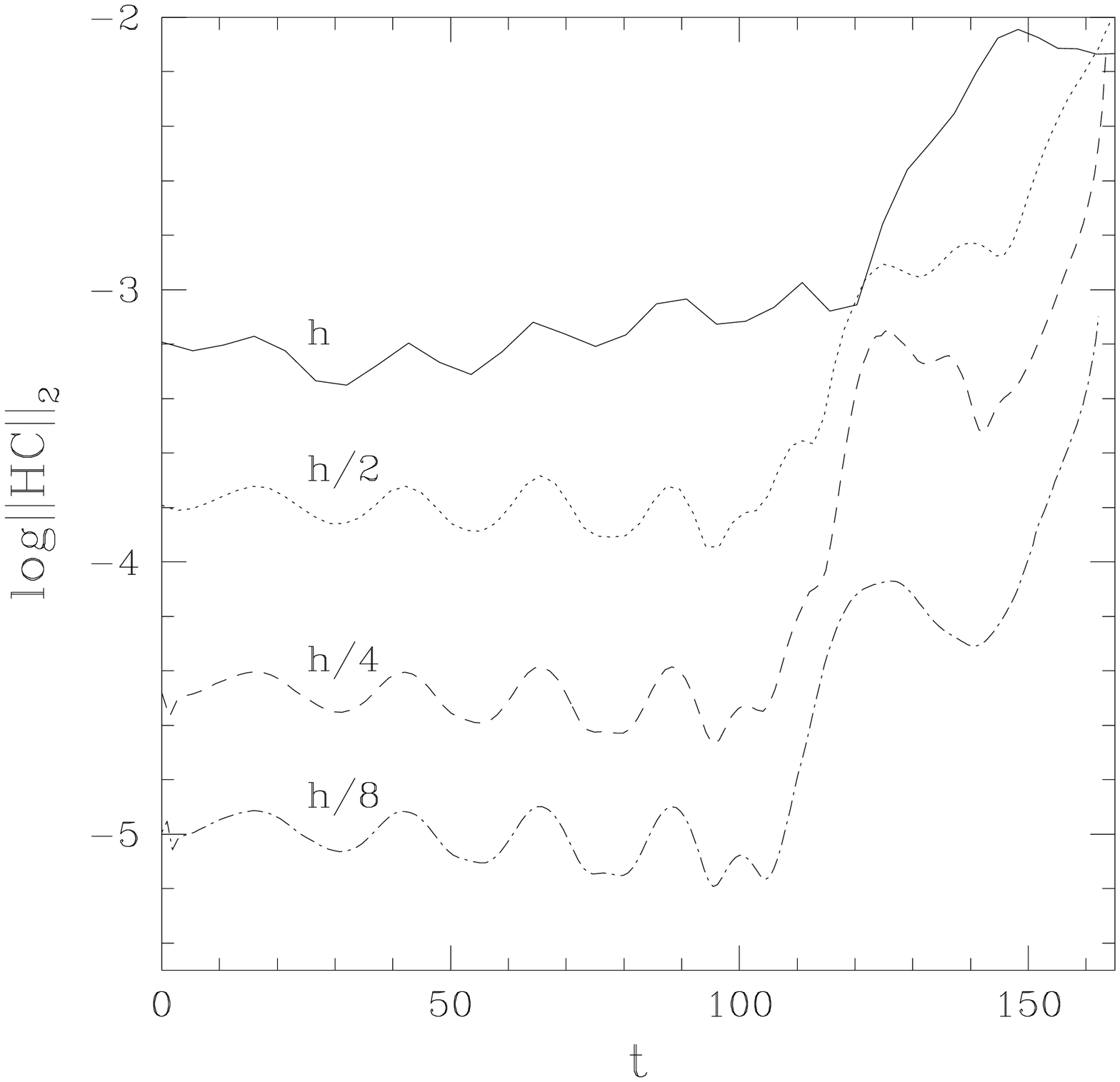}
\end{center}
\caption{The logarithm of the $\ell_2$-norm of the Hamiltonian
constraint as a function of time, evaluated on the portion of the computational domain
lying exterior to the apparent horizon, and from simulations 
at several resolutions of a perturbed black string with $L=1.4L_c$.
As with Fig.~\ref{lambda_1000}, this plot provides evidence that 
convergence is quite good for the $h/2$ and higher resolution 
simulations
(at least until very close to when the supposed coordinate
singularity forms, near $t=165$).
 \label{hc_1000}}
\end{figure}

We now present more detailed results from the simulation of an unstable 
black string evolution.   Specifically, we take 
$L=1.4L_c$, since it is expected that this particular range for the $z$
coordinate will yield something close to the fastest growth rate for the
shortest wavelength instability \cite{gregorylaflame1}.
Because we are now probing uncharted territory with our computations, 
we rely on convergence tests to provide an intrinsic measure of the 
level of error in our calculations.  
To that end, 
we ran the simulation at several resolutions ($n_r \times n_z$): $200 \times 50$
(grid spacing $h$), $400 \times 100$ ($h/2$), $800 \times 200$ ($h/4$), 
and $1600 \times 400$ ($h/8$). 
Due to our use of a compactified radial coordinate, 
the lowest resolution calculation cannot adequately resolve the 
late-time behavior of the solution. However, for the ``medium resolution''
computation with mesh spacing $h/4$, we are apparently within the
convergent regime---see Fig. \ref{lambda_1000}
below for plots of the maximum and minimum areal radii of the apparent horizon as a
function of time, as well as the quantity $\lambda$ defined by
(\ref{lambda}), and  Fig. \ref{hc_1000} for plots of the norm 
of the Hamiltonian constraint as a function of resolution. Therefore, 
unless otherwise noted, all the results shown below are taken from the $h/4$ 
simulation.

\begin{figure}
\begin{center}
\includegraphics[width=15cm,clip=true]{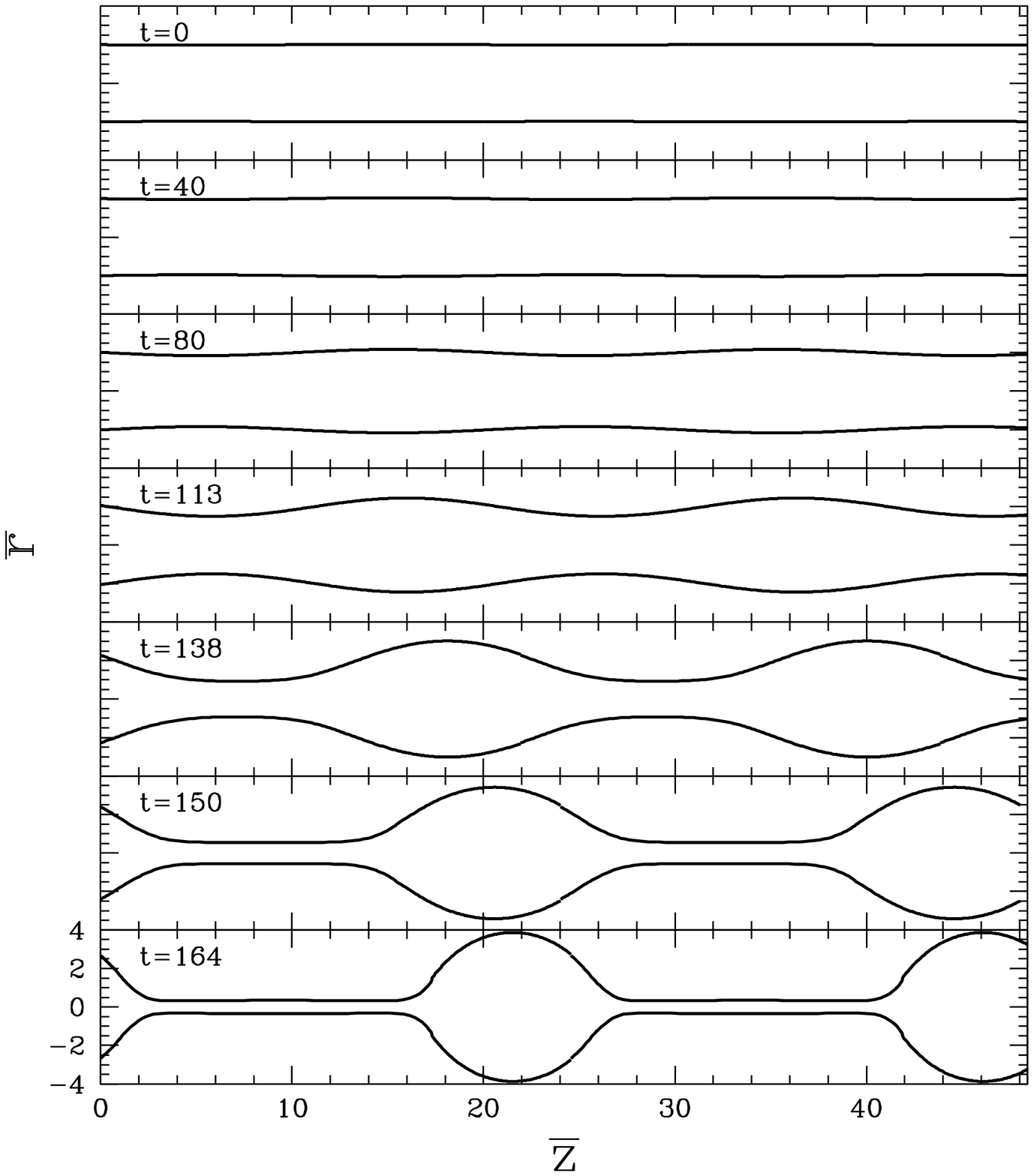}
\end{center}
\caption{Embedding diagrams of the apparent horizon, with the two 
angular dimensions $\theta$ and $\phi$ suppressed, 
from the $h/4$ evolution of a perturbed black string with $L = 1.4 L_c$.
These plots thus describe the intrinsic geometry of the apparent horizon,
at the given instants of constant $t$, in a coordinate system with 
metric $ds^2=d\bar{r}^2 + d\bar{z}^2$. Here, $\bar{z}$ is a periodic coordinate, 
and $\bar{r}$ is the areal radius of $\bar{z}=constant$ sections of the horizon.
To better illustrate the dynamics of the horizon, we have extended the 
solution using the $\bar z$-periodicity, showing roughly two periods of the solution.
See Fig. \ref{ah_embed_1000_length} for a plot of the length of one period
of the apparent horizon versus time.
\label{ah_embed_1000}}
\end{figure}

\begin{figure}
\begin{center}
\includegraphics[width=15cm,clip=true]{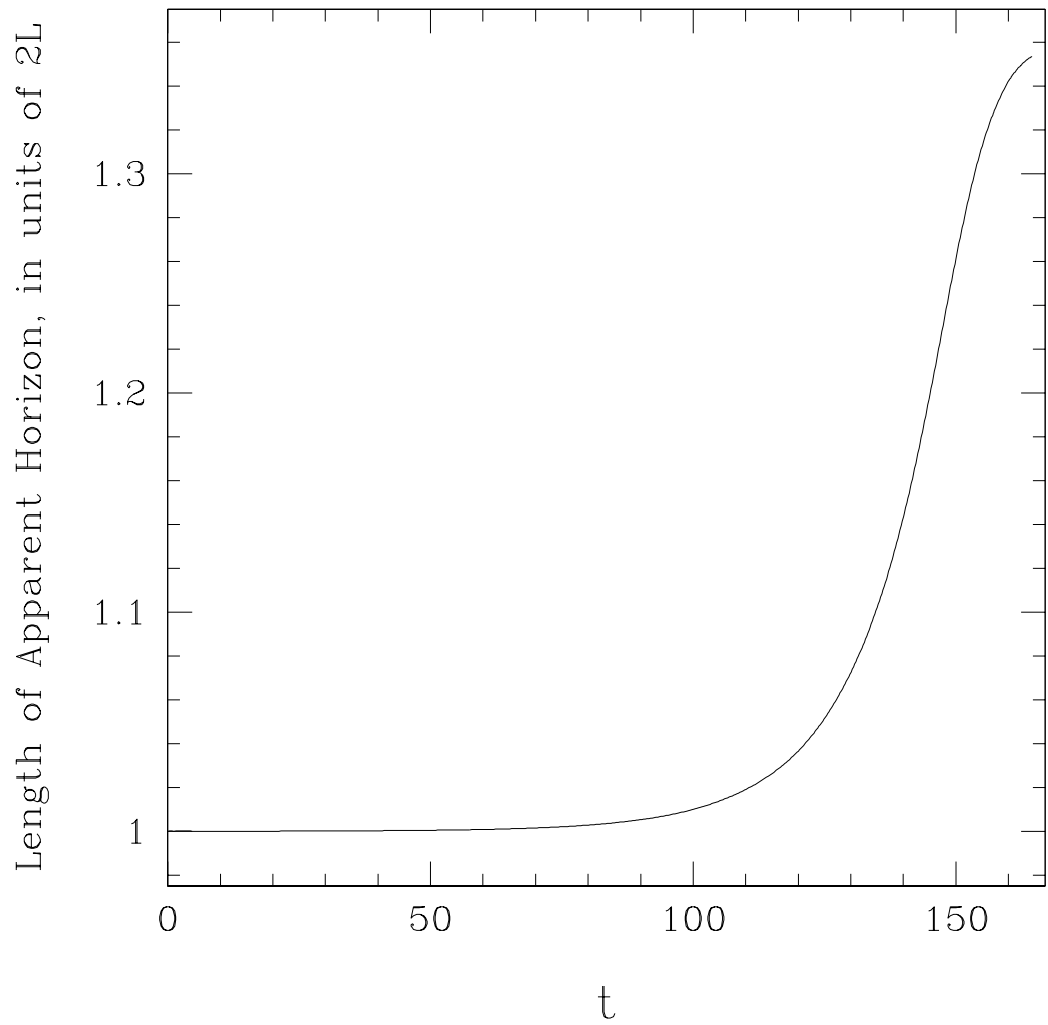}
\end{center}
\caption{The proper length of the apparent horizon curve in the $(r,z)$ plane
(between $z=0$ and $z=L$)
as a function of time, from the $h/4$ evolution of a perturbed black string 
with $L = 1.4 L_c$. 
\label{ah_embed_1000_length}}
\end{figure}

\begin{figure}
\begin{center}
\includegraphics[width=15cm,clip=true]{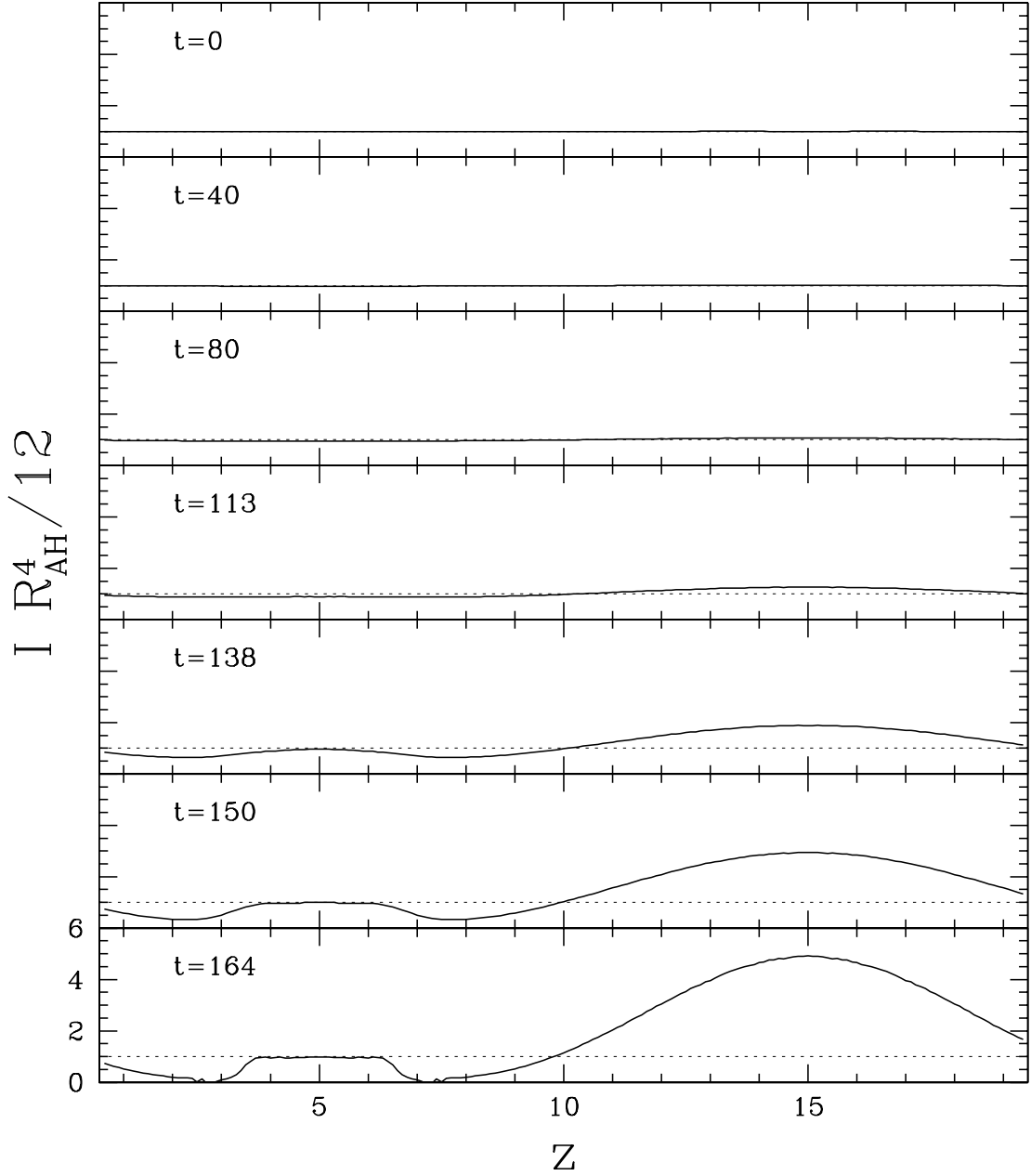}
\end{center}
\caption{The normalized Kretschmann invariant $I^{0N}\equiv I R_{\rm AH}^4/12$ (\ref{I}),
evaluated on the apparent horizon of the 
perturbed black
string spacetime with $L=1.4L_c$  ($h/4$),
at the same times as shown in the embedding diagram
plots (Fig. \ref{ah_embed_1000}). Note however, that here the horizontal axis
is the {\em coordinate} $z$, and in particular the flat region of the
curve between $z\approx3.5$ and $z\approx 6.5$ in the last frame
corresponds to the long, thin neck region shown in the embedding diagram plot.
This demonstrates the rather severe ``grid-stretching'' problems we have then.
For the static black string spacetime, $I^{0N}=1$
(shown for reference as a dotted line in the figure),
while for a static 5D spherical black hole it evaluates to 6. This diagram
therefore also supports the conclusion that at late (simulation) times the 
solution is tending
towards a configuration describable as a sequence of black holes connected
by thin black strings.
\label{I_AH_1000}}
\end{figure}

Fig. \ref{ah_embed_1000} shows embedding diagrams of the apparent
horizon at several times during evolution of the string,
and Fig. \ref{ah_embed_1000_length} shows the proper length of one period
of the apparent horizon (suppressing the angular coordinates) versus time.
(Our embedding uses the vertical axis to represent the areal radius of the apparent
horizon---the horizontal axis is then uniquely determined by requiring that
the length of the curve be equal to the proper length of the horizon).
The simulation crashes shortly after the last time frame shown,
apparently due to the coordinate pathologies that have been 
discussed previously. 
The
embedding diagrams suggest that, at least in the
vicinity of the apparent horizon, the solution is tending towards a 
spacetime that can be
described as a sequence of spherical black holes connected by thin
black strings. 
Additional, quantitative, evidence for this conjecture can be obtained 
through a computation of the curvature invariant, $I$, on the apparent 
horizon. For an exact black string solution, this quantity, which 
we denote $I^0_{\rm BS}$,
is
\begin{equation}
I^0_{\rm BS} = \frac{12}{R_{\rm AH}^4},
\end{equation}
while for the 5-dimensional spherical black hole, the equivalent quantity,
$I^0_{\rm BH}$, is
\begin{equation}
I^0_{\rm BH} = \frac{72}{R_{\rm AH}^4},
\end{equation}
where, in both of the above expressions,
$R_{\rm AH}$ is the areal radius of the horizon. In 
Fig. \ref{I_AH_1000} we plot the normalized quantity 
\begin{equation}
I^{0N}\equiv \frac{I}{I^0_{\rm BS}} = \frac{I R_{\rm AH}^4}{12}
\end{equation}
evaluated on the apparent horizon of our
numerical solution of the unstable spacetime---$I^{0N}$ is $1$
for a black string, and $6$ for a black hole. The figure shows that,
as judged by $I^{0N}$, 
the part of the apparent horizon that is forming a long neck always resembles
a black string---the part that is forming
a bulge, however,  has a value of $I^{0N}$ tending towards that 
corresponding to a black hole.
At $R_{{\rm max}}$, $I^{0N}$ has only reached $\sim 5$ by the time the 
simulation ends; however, the behavior 
of $R_{{\rm max}}$ seen in Fig.
\ref{lambda_1000} suggests that the growth in the normalized curvature 
invariant, though slowing down, should continue.
Fig. \ref{I_AH_1000} also demonstrates the grid-stretching problems
that we surmise are causing the code to eventually crash---in that plot we 
use the coordinate $z$ as the horizontal axis, and observe that the 
relatively small region where $I^{0N}\approx1$ corresponds to the long 
neck in Fig. \ref{ah_embed_1000}. 
In particular, in the vicinity of the ``neck'', $g_{zz}$ becomes
quite large, as do its derivatives. 

\begin{figure}
\begin{center}
\includegraphics[width=20cm,clip=true]{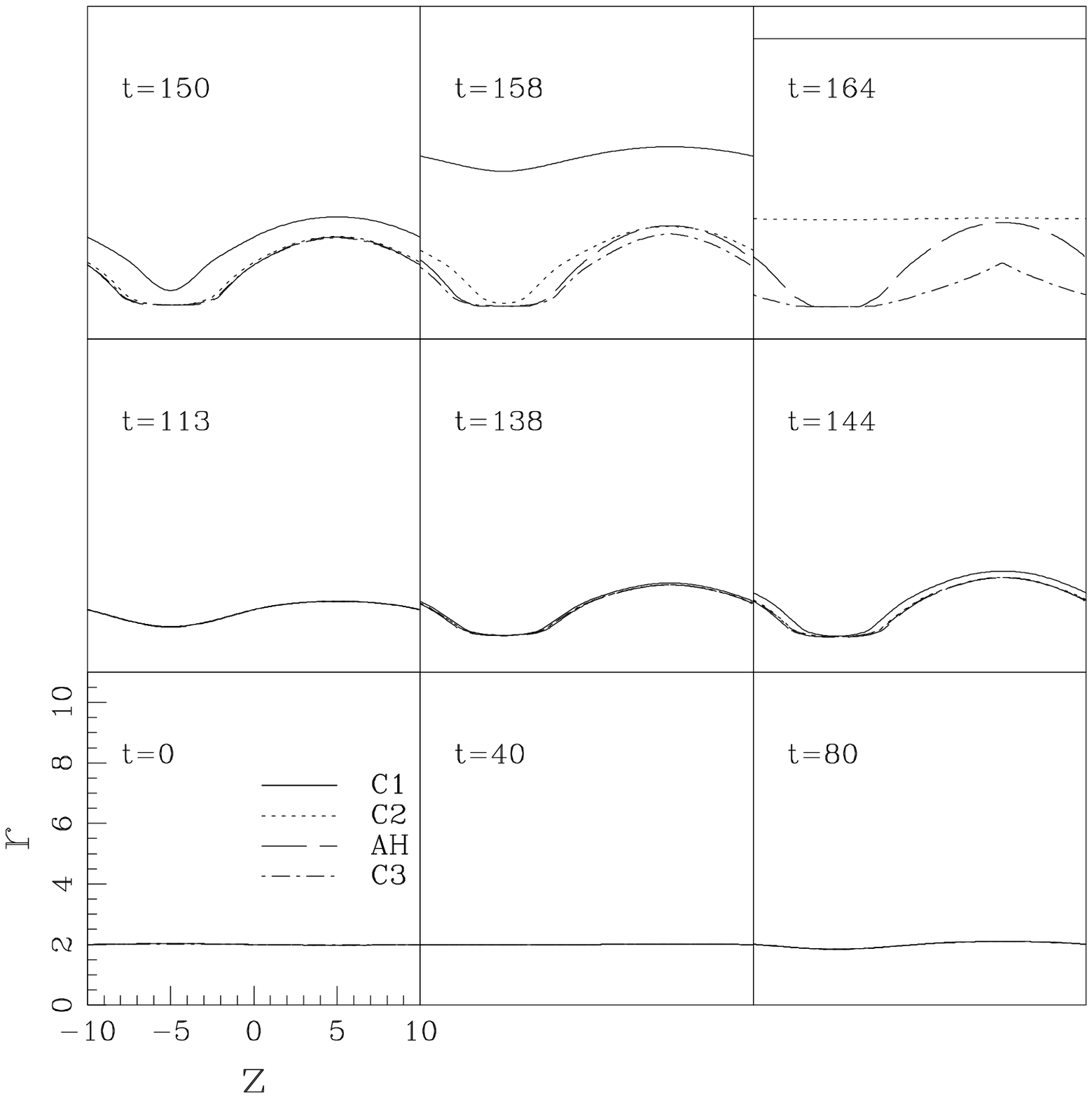}
\end{center}
\caption{Plots of the apparent horizon (labeled AH) and estimates
of the event horizon location (C1, C2 and C3)
in coordinate space (in contrast to the embedding coordinates used in
Fig. \ref{ah_embed_1000}), from the evolution of a
perturbed black string with $L=1.4L_c$, computed at resolution
$h/4$. Here, the C1 (C2) curve marks the inward-directed past light-cone
of the surface $r=10$ ($r=4$) at $t=164$. C3 denotes the outward-directed past of a
surface just inside the apparent horizon at $t=164$. Thus, moving backwards in time,
these curves should asymptote towards the event horizon of the spacetime. These
plots suggest that for most of the evolution (at least), the apparent horizon 
is an excellent approximation to the event horizon.
\label{ah_eh}}
\end{figure}

Finally, in Fig. \ref{ah_eh} we show plots of the approximate event
horizon (as described in Sec. \ref{sec_monitor}), together with the apparent horizon 
for the simulation. The results shown in the plot suggest that our computed
apparent horizon is an excellent approximation to the event horizon, at 
least at early times
(not much can be said regarding the late time behavior of the event horizon, as the spacetime
has not settled down to a stationary state when the simulation ends).

\section{Conclusions}\label{sec:conclusion}
We have performed a preliminary numerical study of the
instability of the 5-dimensional black string. Coordinate
pathologies prevent us from definitively identifying
the final end-state(s) of an unstable black string. This claim is supported
by the fact that the code crashes at very nearly the same time at varying
resolution, and that curvature invariants remain well behaved throughout
the evolution. The former suggests that a numerical instability is not 
responsible for the crash, while the latter indicates that a physical 
singularity is probably also not to blame.
Despite the premature termination of the simulation, we 
find evidence that
the spacetime evolves towards a configuration that
looks like a sequence of black holes connected by thin black
strings, and characterized by an expansion of the string direction. 
Since the spacetime is still 
fairly dynamical at the time our 
simulations end, we cannot deduce how close this state is to a final
configuration. Nevertheless, the dynamical behavior
observed is sufficiently robust for some comments to be made. For instance, 
the results are not
inconsistent with Gregory and Laflamme's conjecture that the
solution bifurcates into a sequence of black holes---indeed, we can suggest
at least two mechanisms by which this could occur:
\begin{enumerate}
\item
Via a thinning neck that eventually vanishes, if the
trend seen in the simulation continues. Note that this would
require a) that the proper length of the string continue to grow, in order 
not to violate area theorems, and b) that the thin string be non-uniform,
for otherwise
{\em it} would be subject to a further Gregory-Laflamme-like instability.
\item
Via a sequence of Gregory-Laflamme instabilities, if the thin neck stays
``close'' to a uniform black string, since the neck's length is 
beyond the critical one for a string with an effective mass computed from 
the radius of the apparent horizon.
In this case, one could
envision a ``cascade'' of instabilities leading to the bifurcation. 
\end{enumerate}

We note that either scenario would not necessarily be inconsistent with Horowitz
and Maeda's results, should the vanishing of the
neck take infinite affine time as measured by local null
generators of the horizon.  At the same time, a continuation of the 
observed trend 
would argue against achieving a stationary solution with a mild dependence
on the string dimension (i.e. a small value of $\lambda$), as found in
perturbative calculations \cite{gubser}. For then, the rather
extreme thinning/bulging that we see must be transient behavior that is
``further'' from the end-state than the perturbed black string was.

A more complete exposition of the nature of unstable black string 
evolution would appear to require coordinate conditions able to adapt to
solution features that develop at late times---that is, in a manner that 
does not introduce  severe metric gradients that are not
correlated with large gradients in physical quantities.  
For example, it may help to
replace the fixed-lapse slicing with maximal slicing, which
enforces that the divergence of the local, spatial volume element
be zero. Another, perhaps even more crucial option, would be to
introduce a $z$ component to the shift vector that keeps $g_{zz}$
close to (or exactly) unity throughout the evolution. These options
are currently under investigation.

Additionally, it would be interesting to explore a wider range of initial
conditions describing ``perturbed'' black strings than that considered
here. For example, the imposed
$z$-periodicity implies that the equivalent, uncompactified space
time consists of identical spherical-black-hole/black-string
segments at late (intermediate) times. It would be instructive to
see what happens should we break this symmetry, by making $L \gg L_c$,
and then introducing some higher-wavelength perturbation similar to
$q>1$ in (\ref{gtt}), but with more asymmetry in the initial data (note that
this would be more computationally demanding).
Finally, it would be very interesting to study the evolution of the 
solutions recently found by Wiseman~\cite{wiseman}, and mentioned in the 
introduction.
These configurations actually correspond to stationary
solutions, and their perturbative stability, or otherwise, is currently not 
known.
Since Wiseman shows that his solutions cannot be the end-states
conjectured by Horowitz and Maeda, it is important to
understand their behavior, since if they are stable they may well 
represent physically meaningful states, while if unstable, they may
be difficult to attain via dynamical evolution. An interesting observation
from our simulations, to the length so far achieved, is that they do not
display a conical ``waist'' like those
presented in~\cite{wiseman} and further analyzed in~\cite{wisemankol}.

\section{Acknowledgments} We gratefully acknowledge support from the 
following agencies, institutes and grants:  NSERC, NSF PHY-0099568, The 
Canadian Institute for Advanced Research,
The Canadian Institute for Theoretical Astrophysics, The Pacific Institute for 
Mathematical Sciences,  The Government of the Basque Country,
The Izaak Walton Killam Fund and Caltech's Richard Chase Tolman Fund.
Computations were performed on
(i) the {\tt vn.physics.ubc.ca} cluster which was funded by the Canadian Foundation for
Innovation (CFI) and the BC Knowledge Development Fund; (ii) {\tt LosLobos} at 
Albuquerque High Performance Computing Center 
(iii) The high-performance computing facilities within LSU's Center for
Applied Information Technology and Learning, which is funded through
Louisiana legislative appropriations, and (iv) The {\tt MACI} cluster
at the University of Calgary, which is funded by the Universities of Alberta,
Calgary, Lethbridge and Manitoba, and by C3.ca, the Netera Alliance, CANARIE, 
the Alberta Science and Research Authority, and the CFI.
We would like to thank G. Horowitz, W.G. Unruh, R.
Wald, R. Myers, T. Wiseman and B. Kol for stimulating discussions.

\begin{appendix}

\section{Initial data solver}\label{app:initialdata}
We solve the set of coupled constraint equations
(\ref{HC})-(\ref{MC}) via an iterative procedure, where at each sub-step
of the iteration we solve a single equation for one of 
$g_{rr}$, $k_{\theta\theta}$ or $k_{rr}$, assuming that the values 
of the other variables are known.
We iterate this process until the residuals of all the equations are 
simultaneously below a certain tolerance---a typical value is $10^{-5}$. 
The overall iteration is initialized using values corresponding to 
an unperturbed black string solution. We now provide a few more details 
concerning the solution of each of the constraint equations.

The equations are discretized on a uniform grid of points ($x_i,
z_j$) with $i=1,...,N_x$, and $j=1,...,N_z$ (recall that, from~(\ref{defx}),
the
radial coordinate, $r$, is related to $x$ by $r=x/(1-x)$). 
We first consider the Hamiltonian constraint 
(\ref{HC}) which, in the coordinate system we have adopted,
has the following form:
\begin{equation}
\label{hc} F_1             \frac{\partial   g_{rr}}{\partial x}
+ F_2 g_{rr} \frac{\partial^2 g_{rr}}{\partial z^2} + F_3 g_{rr}
\frac{\partial   g_{rr}}{\partial z}   + F_4
\left(\frac{\partial   g_{rr}}{\partial z}\right)^2 + F_5
\left(g_{rr}\right)^2 + F_6       g_{rr} =0 \, .
\end{equation}
Here, the $F_m, \, m = 1, \ldots, 6$, are functions that generally depend 
on all the metric
coefficient and their derivatives {\em except} $g_{rr}$ (and its
derivatives). We discretize this equation to second order
in the mesh spacing using a difference approximation centered at 
the points $(x_{i+1/2}, z_j)$.
Because of the discretization used and the form of
(\ref{hc}), we can solve the resulting set of equations ``line-by-line'' 
in $x$, starting at the inner boundary, $i=1$, which is chosen well
within the horizon of the string (typically at $r=M$), and where 
the boundary values, $[g_{rr}]_{1,j}, \, j = 1,\ldots N_z$, are those 
corresponding to an unperturbed black string. As we integrate outwards 
in $x$, the determination of the $i$-th line of unknowns,  
$[g_{rr}]_{i,j}, \, j = 1,\ldots N_z$, involves the solution of 
a $N_z$-dimensional, non-linear, cyclic (because of the $z$-periodicity), tridiagonal 
system that we solve using Newton's method and a cyclic tridiagonal 
linear solver~\cite{nr}.

We now direct attention to the $r$
momentum constraint, which, viewed as an equation for $k_{\theta \theta}$,
has the form:
\begin{equation}
G_1 \frac{\partial k_{\theta \theta}}{\partial x} + G_2 k_{\theta
\theta} + G_3 =0,
\end{equation}
where the $G_m, \, m = 1,2,3$ do not depend on $k_{\theta \theta}$ or its
derivatives.  We note that there is complete decoupling in the $z$-direction
in this case; in effect, we have to solve an ODE along each $z={\rm const.}$ 
line.  We again discretize using second-order finite difference techniques,
fix the  boundary values $[k_{\theta\theta}]_{N_x,j}, \, j = 1,\ldots N_z$,
at $x=1$ ($i_o$) using the unperturbed
black string solution, then solve for the remaining unknowns, marching 
inwards in $x$.

Finally, the $z$ component of the momentum constraint, which fixes 
$k_{rr}$, has the form:
\begin{equation}
H_1 \frac{\partial k_{rr}}{\partial z} + H_2 k_{rr}  + H_3 =0 \, ;
\end{equation}
where the $H_m, \, m = 1, 2, 3$ are independent of $k_{rr}$ and its 
derivatives.  This equation is solved analogously to the $r$ momentum
constraint, but now using a discretization that is centered at points ($x_i,z_{j+1/2}$).
``Boundary values'', $[k_{rr}]_{i,1}, \, i = 1,\ldots N_x$, are specified along 
the line $z=z_{{\rm min}}$, again using corresponding values from the
black string solution, and the integration proceeds for $j = 2, 3, \ldots N_z$.

\section{Finding apparent horizons}\label{app:AH}
We use a {\em flow}, or {\em level-set} method to search for
apparent horizons within $t={\rm const.}$ spatial slices of
the spacetime.  We restrict our search to simply connected apparent horizons
that are periodic in z. Such an apparent horizon can be described by a curve in
the $(r,z)$ plane, which we define to be the level surface $F=0$
of the function
\begin{equation}\label{fs2}
F(r,z) = r - R(z).
\end{equation}
In other words, the apparent horizon will be given by the curve $r=R(z)$. The 
apparent horizon
is the outermost, marginally trapped surface; hence, we want to
find an equation for $R(z)$ such that the outward null expansion,
normal to the corresponding surface $F=0$, is zero. To this end,
we first construct the unit spatial vector $s^a$, normal to
$F={\rm const.}$:
\begin{equation}\label{sadef}
s^a=\frac{g^{ab} F_{,b}}
          {\sqrt{g^{cd} F_{,c}F_{,d}}}.
\end{equation}
Then, using $s^a$ and the $t={\rm const.}$ hypersurface normal vector $n^a$,
we can construct future-pointing outgoing($+$) and ingoing($-$) null
vectors
\begin{equation}\label{null}
\ell^a \pm=n^a \pm s^a.
\end{equation}
The normalization of the null vectors is (arbitrarily)
$\ell^a_+ \ell_{-a}=-2$.
The outward null expansion $\theta_+$ is then the divergence of $\ell^a_+$
projected onto an $F={\rm const.}$ surface:
\begin{equation}\label{exp1}
\theta_+ = \left(g^{ab}-s^a s^b \right) \nabla_b \ell_{+a}.
\end{equation}
Substituting expressions (\ref{fs2}) and (\ref{sadef}) into equation (\ref{exp1}),
with $\theta_+=0$, provides us with an ordinary differential equation
for $R(z)$.

Initially we solved equation (\ref{exp1}) via a ``shooting''
method---given a guess for $R(0)$, and assuming that
$R'(0)=0$ (where a prime denotes differentiation with respect to $z$), 
we integrate the equation to $z=z_L$, and repeat
the process until we find a solution where $R(z_L)=R(z)$ and
$R'(z_L)=0$. An efficient sequence of guesses can be
generated using a bisection search, as the qualitative behavior of
the solution is different depending upon whether the initial guess
for $R(0)$ is inside or outside the apparent horizon.

Since the shooting method is difficult  to extend to a parallel
implementation in an efficient way, we opted to use the
following point-wise relaxation method (or flow method) to determine
$R(z)$. We supply an initial guess, $R_0(z)$, for the entire
function $R(z)$, and then iterate the following equation until the
norm of the expansion $\theta_+(z)$ of $F(0)$ is below some
desired threshold (in our runs we have typically set it to
$10^{-3} h$, where $h$ is the basic scale of discretization):
\begin{equation} \label{flow_eqn}
\Delta R_n(z) \equiv R_{n+1}(z) - R_{n}(z) = - \theta_+(z) \Delta
\tau \, .
\end{equation}
Here, $R_{n}(z)$ is the solution after the $n^{th}$ iteration.
Equation~(\ref{flow_eqn}) can be viewed as an {\em explicit}
discretization of a parabolic evolution equation for
$R(z,\tau)$, where $\tau$ is ``time'' and $\Delta \tau$ is the
time-step for the evolution (the parabolic nature of the equation
is evident when $\theta_+$ is expanded in terms of $R(z)$ via
(\ref{exp1})---for brevity we do not give the explicit form of the
equation here). Thus, for stability $\Delta \tau$ must be chosen to
be less than $\Delta z^2$.

Given a ``reasonable'' initial guess $R_0(z)$, one can see how
iteration of equation (\ref{flow_eqn}) will cause $R_n(z)$ to ``flow''
to the apparent horizon: if $R_n(z)$ is outside of the apparent horizon, then typically the
expansion $\theta_+(z)$ will be positive there, causing
$R_{n+1}(z)$ to decrease towards the apparent horizon, and vice-versa if
$R_n(z)$ is inside the apparent horizon. We use $R(z)=2$ as the
initial guess at $t=0$; after $t=0$ we search for the apparent horizon every $N$
time steps (where $N$ is typically in the range $10-30$), and use
the shape found at the previous search as the initial guess for
the next search. Usually, on the order of tens to thousands of
iterations of (\ref{flow_eqn}) are required to solve for $R(z)$ to
within a level of accuracy such that the approximate solution is
roughly within a mesh spacing of the exact solution (as estimated
in a few specific calculations by solving (\ref{flow_eqn}) close to
machine precision). A single iteration of (\ref{flow_eqn}) can be
computed very rapidly relative to the time taken to compute a
metric-evolution step; however in a parallel environment, if
thousands of iterations are needed on a regular basis (which {\em
is} so at late times during the evolution of an unstable black
string), the apparent horizon finder becomes a slight speed bottleneck in
the code, due to the time it takes to communicate the results of
each iteration
amongst the processors involved.

\section{Finding (Approximate) Event Horizons}\label{app:null}
Here we describe one method we use, following~\cite{eh}, to locate 
approximations to event horizons. This method involves
locating the boundary of the causal past of some $r={\rm const.}$ surface
of the spacetime by following radial null geodesics.

We write the geodesic equation in Lagrangian form:
\begin{equation}
{\mathcal L} =     g_{tt} \left(t^\prime\right)^2
               + 2 g_{tr} t^\prime r^\prime
               +   g_{rr} \left(r^\prime\right)^2
               + 2 g_{rz} r^\prime z^\prime
               +   g_{zz} \left(z^\prime\right)^2,
\end{equation}
where $\lambda$ is the affine parameter along the geodesic, and a
prime denotes differentiation with respect to $\lambda$. Since we are
interested in null trajectories, we set ${\mathcal L}=0$. For radial,
ingoing geodesics, we have $\theta^\prime=\phi^\prime=0$, and thus the geodesic
equations reduce to the set:
\begin{eqnarray}
\label{eq:geod}
\dot r & = & \frac{\alpha}{\sqrt{g_{rr}}} - \beta, \\
\dot \lambda & = & \frac{2 \alpha \sqrt{g_{rr}}}{\Pi_r}, \nonumber
\end{eqnarray}
where the dot denotes a derivative with respect to coordinate time, and
$\Pi_r={\partial{\mathcal L}}/{\partial r^\prime}$. Then, starting at a certain
value of $r=r_0$, and for each grid point along the $z$ direction, equations
(\ref{eq:geod}) are integrated
backwards in time using a second order Runge-Kutta scheme.

Following null geodesics along $z={\rm const.}$ lines does not guarantee that we
are tracing the causal past of $r=r_0$, though for the spacetimes studied here,
and the coordinate system used, this should offer a good approximation. Furthermore,
although radial geodesics might not be the best choice, since the event horizon is an
attractor, they will trace it accurately. (For related discussions of
approximate event horizon location in the axisymmetric four-dimensional case, 
see~\cite{shapiroteukolskyEH,seidelEVENTHORIZONS,caveny}.)

\end{appendix}

\end{document}